\documentclass[11pt]{article}
\usepackage[preprint]{acl}
\author{Yuru Liang\textsuperscript{1,2}\thanks{Co-first authors. Work done as \mbox{DigitForce} interns.}\quad
  Guangyinan Mei\textsuperscript{1,2}\footnotemark[1]\quad
  Miles Fan\textsuperscript{1}\thanks{Corresponding author.}\\
  \normalfont\textsuperscript{1}Renmin University of China\qquad\textsuperscript{2}DigitForce\\
  \normalfont\small\texttt{yuru@ruc.edu.cn}\quad
  \texttt{mgyn1.art@gmail.com}\quad
  \texttt{fmh4us@gmail.com}}
\newcommand{\codeFootnote}{\footnote{Code and experiment scripts: \url{https://github.com/fmh1art/AskWise}.}}

\usepackage{microtype}
\AtBeginDocument{%
  \raggedbottom
  \setlength{\floatsep}{8pt plus 2pt minus 2pt}%
  \setlength{\textfloatsep}{12pt plus 2pt minus 2pt}%
  \setlength{\intextsep}{9pt plus 2pt minus 2pt}%
  \setlength{\dblfloatsep}{9pt plus 2pt minus 2pt}%
  \setlength{\dbltextfloatsep}{12pt plus 2pt minus 2pt}%
}

\usepackage{times}
\usepackage{latexsym}
\usepackage[T1]{fontenc}
\usepackage[utf8]{inputenc}
\usepackage{amsmath,amssymb,booktabs,graphicx}
\newcommand{\method}{\textsc{AskWise}}
\title{AskWise: Execution-Guided Clarification of Intent and Data\\in Text-to-SQL}
\begin{document}
\maketitle
\begin{abstract}
Two interpretations of a database question can agree on the observed records and diverge after a correction. A single record can therefore determine whether the user needs to clarify their intent.
\method{} executes candidate queries across possible repairs. It asks about a choice only when two executions differ in that choice alone and give different answers. Once all surviving combinations give the same known answer, it returns that answer.
For Boolean answers on a uniform binary cube, we derive an exact bound on the difference in EC2 gain between two questions from partial execution results.
Across 120 sales tasks spanning three query types, the bound reduces mean executions per episode from 14.88 to 13.80 relative to a partition relaxation. Asking at the first sign that a choice matters uses 13.18 executions and 0.025 more questions per episode.
On AmbiQT, a controlled value swap separates the reference queries in 424 of 1,206 eligible cases. On 155 sampled questions, pruning the same model candidates across both database states raises recall of reference answer signatures by 8.71 percentage points for Doubao Lite and 10.00 for GPT Sol.
\end{abstract}
\section{Motivation And Contributions}
A database assistant can ask what a user means or whether a record is correct.
Interactive semantic parsers ask users to resolve competing query interpretations \citep{gur-etal-2018-dialsql,yao2019misp,elgohary2020feedback,tian-etal-2023-interactive}.
A user's name for an entity may also differ from its stored value \citep{pham2026raqs}.
Query-driven cleaning identifies records worth checking for a given query \citep{Luyi_Mo_2013,altwaijry2013query}.
\emph{When both the query and the data are uncertain, how can an assistant tell what it still needs to ask?}

\begin{figure}[t!]
\centering
\includegraphics[width=\columnwidth]{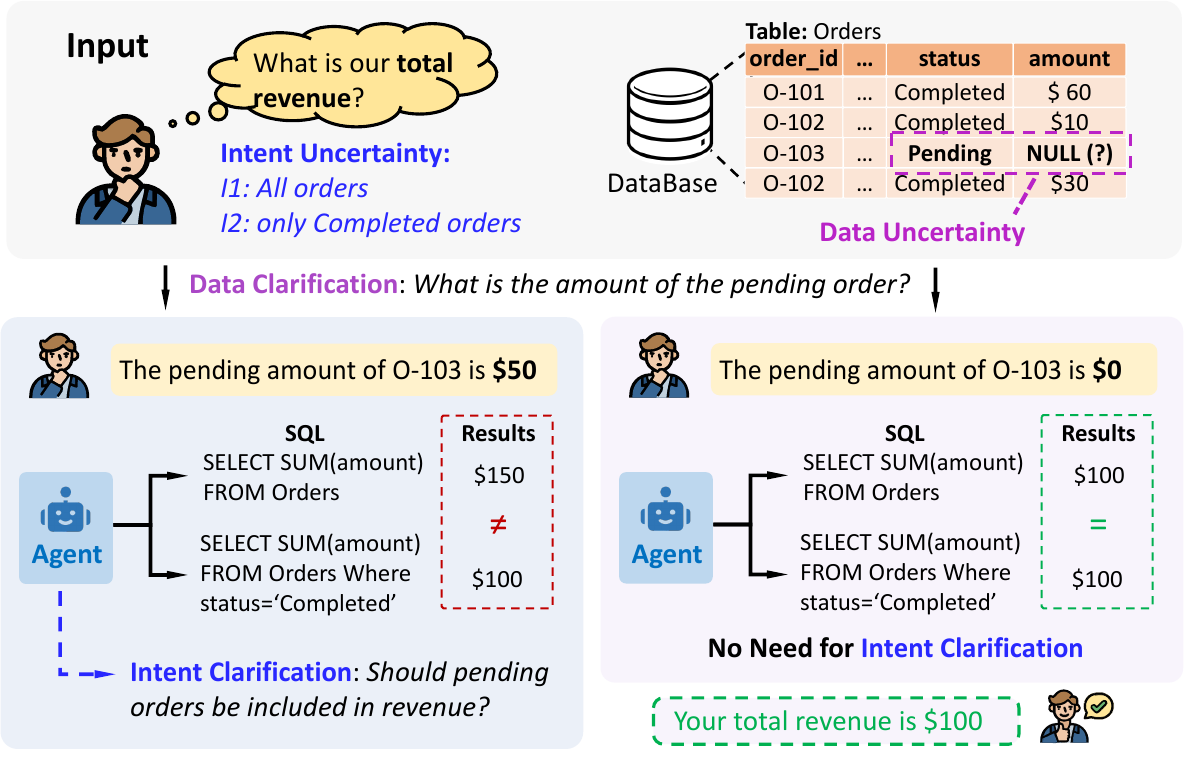}
\par\vspace{-3pt}
\caption{Clarifying the amount for pending order \texttt{O-103}. At \$50, including it raises revenue from \$100 to \$150 (left). At \$0, both queries return \$100 (right).}
\label{fig:overview}
\vspace{-3pt}
\end{figure}

The user in Figure~\ref{fig:overview} asks for total revenue.
Three completed orders are worth \$60, \$10, and \$30. A pending order has no amount on record.
SQL's \texttt{SUM} skips the missing value, so summing all orders and summing only completed orders both give \$100.
If the pending amount is \$50, the two totals become \$150 and \$100. The assistant now needs to know whether pending orders count.
If the amount is zero, either reading gives \$100. If the user chooses completed orders first, the total is \$100 regardless of the missing amount.

\begin{figure*}[t]
\centering
\includegraphics[width=\textwidth]{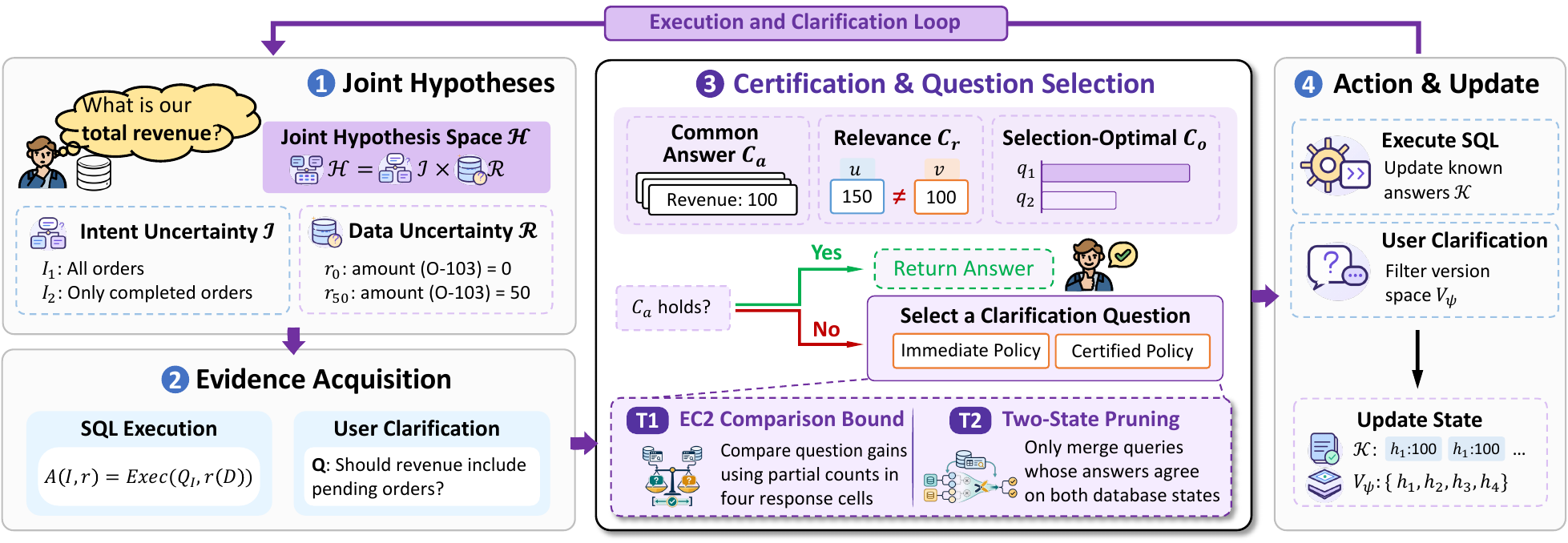}
\par\vspace{-3pt}
\caption{One round of \method{}. T1 compares question gains from partial executions. T2 merges queries only if their answers agree on both database states.}
\label{fig:policy}
\vspace{-3pt}
\end{figure*}

One approach merges queries with identical observed answers \citep{yin2026air}. Here that would conceal their disagreement after a \$50 repair. Asking about every missing value would make users investigate orders they mean to exclude.
In our audit of AmbiQT \citep{bhaskar-etal-2023-benchmarking}, we swapped two values in one column and reran both reference queries. Of 1,206 eligible pairs, 424 agreed before the swap and differed afterward.

\method{} uses SQL executions to test which choices matter (Figure~\ref{fig:policy}). Each hypothesis pairs a query interpretation with a repaired database.
Two surviving hypotheses that differ only in one choice and return different answers give a \emph{relevance certificate} for asking about that choice.
Once every surviving hypothesis has produced the same result, the assistant has an \emph{answer certificate} and can stop while some choices are still unresolved.

Several questions may have a relevance witness. Equivalence class edge cutting (EC2) ranks them by the expected weight of edges they eliminate between hypotheses with different answers \citep{golovin2010ec2}.
Two runs can show that intent matters while leaving open whether asking about the record has greater gain. Each further execution supplies one missing answer. For Boolean answers on a uniform binary cube, two coordinate questions divide the worlds into four cells. The observed positives and the number of unexecuted worlds bound the final positive count in each cell. From these counts we bound the gain difference and can certify a maximizer before all worlds are executed.
One policy waits for this \emph{action certificate}. The other asks at the first witness.

We make three contributions.
\begin{itemize}
\item We formulate clarification over a finite set of interpretations and repairs, and give conditions for certifying both a question's relevance and the returned answer (Section~\ref{sec:problem}).
\item We express the difference in EC2 gain between two coordinate questions using four counts. This identity gives attainable bounds on the difference before all Boolean answers are known (Section~\ref{sec:certificates}).
\item On the 120 Sales Test tasks with fixed execution order, certifying a maximizer takes 0.62 more executions and saves 0.025 questions per episode relative to asking at the first witness. On AmbiQT, we measure the loss in reference coverage from observed-state pruning and inspect 100 executable queries that match neither reference (Section~\ref{sec:experiments}).\codeFootnote
\end{itemize}

\section{Related Work}
\paragraph{Interactive Parsing.}
NaLIR resolves query ambiguity through dialogue \citep{li2014nalir}.
DialSQL asks users to validate parts of a SQL query \citep{gur-etal-2018-dialsql}, and MISP selects when and where to ask \citep{yao2019misp}.
Users can also revise a parse through natural-language corrections \citep{elgohary2020feedback} or by editing a stepwise explanation \citep{tian-etal-2023-interactive}.
SParC studies follow-up questions, and CoSQL includes clarification and unanswerable requests \citep{yu-etal-2019-sparc,yu-etal-2019-cosql}.
\method{} asks whether a missing value makes an intent question necessary.

\paragraph{Clarification Questions.}
AmbigQA pairs alternative answers with question rewrites \citep{min2020ambigqa}.
AmbiQT pairs competing SQL queries \citep{bhaskar-etal-2023-benchmarking}, and AMBROSIA tests scope, attachment, and vagueness \citep{saparina2024ambrosia}. \citet{saparina-lapata-2025-disambiguate} generate missing interpretations before parsing. PRACTIQ studies dialogues about ambiguous and unanswerable requests \citep{dong-etal-2025-practiq}.
CLAQUA and CAmbigNQ evaluate when to ask, what to ask, and how to use replies \citep{xu-etal-2019-asking,lee-etal-2023-asking}.
CLAMBER tests whether models recognize and clarify ambiguity \citep{zhang-etal-2024-clamber}.
\citet{wang-etal-2023-know} locate ambiguity in SQL requests. CLARITY tests how parsers handle partial replies \citep{sarwar-etal-2026-clarity}.

\paragraph{Query Equivalence.}
AIR removes interpretations with identical outputs on the observed data \citep{yin2026air}.
SOMA-SQL uses synthetic logs and execution probes \citep{somayajula2026soma}, and CircleClar studies which answers a model can reach through repeated clarification \citep{tang2026circleclar}.
Executing queries on additional databases can expose differences that a single instance conceals \citep{zhong2020testsuite}.
Cosette proves equivalence or finds counterexamples for supported SQL fragments \citep{chu2017cosette}.
VeriEQL verifies bounded equivalence under integrity constraints \citep{he2024verieql}, and SpotIt+ searches for distinguishing databases under mined constraints \citep{tremante2026spotitplus}.
For clarification, a counterexample must distinguish the queries on a repair consistent with the user's replies.

\paragraph{Data Cleaning.}
Selective cleaning for top-$k$ queries and query-driven entity resolution target records that affect a query result \citep{Luyi_Mo_2013,altwaijry2013query}.
QOCO consults experts about wrong or missing answers and the records responsible \citep{bergman2015qoco}. ActiveClean selects records to clean while updating a statistical model \citep{krishnan2016activeclean}.
HoloClean infers repairs from constraints, external information, and statistical signals \citep{rekatsinas2017holoclean}.
Consistent query answering finds individual answers shared by all admissible minimal repairs \citep{arenas1999consistent}.
We test whether the full output agrees across candidate repairs and interpretations.
Semiring provenance tracks dependencies on annotated records \citep{green2007provenance}.
We group worlds by bounded integer score using generating functions and differential inference \citep{todor2013massive,darwiche2003differential}.

\paragraph{Adaptive Selection.}
Generalized binary search selects questions to identify a hidden hypothesis \citep{nowak2011geometry}.
EC2 targets its answer class \citep{golovin2010ec2}, and decision-region determination allows overlapping sets of acceptable answers \citep{javdani2014decision}.
Adaptive submodularity gives conditions for greedy guarantees and lazy evaluation \citep{golovin2011adaptive}.
Stochastic Boolean function evaluation and score classification study the cost of revealing inputs sequentially \citep{deshpande2016stochastic,gkenosis2018score}.
EC2 takes the answer classes as input. Several worlds can share an answer. Here the policy must execute candidate queries on the proposed repairs to learn which worlds belong to each class. An execution labels one world and leaves the others admissible. A user response can rule out several worlds whose answers are still unknown. The policy can execute another world or ask about a choice with a known witness.

\section{Clarification Model}
\label{sec:problem}
\subsection{Joint Hypotheses}
Let $D$ be the database, $\mathcal I$ a finite set of interpretations, and $\mathcal R$ a finite set of repairs.
A \emph{world} pairs a query $Q_i$ with a repaired database $r(D)$.
Its answer is defined for each admissible pair $(i,r)\in\mathcal H\subseteq\mathcal I\times\mathcal R$ by
\begin{equation}
 A(i,r)=\operatorname{Exec}(Q_i,r(D)).
\end{equation}
In the SQL audit, we compare outputs as multisets of typed rows.
This preserves duplicate rows and ignores row order. Column order within each row still matters (Section~\ref{sec:sql-errors}).
Two queries can give the same output on a particular database even when they differ on other instances \citep{zhong2020testsuite,chu2017cosette}.

A clarification question $q$ reveals $q(h^\star)$ for the hidden world $h^\star$. After responses $\psi$, the surviving worlds form the version space
\begin{equation}
 V_\psi=\{h\in\mathcal H:\forall q\in\operatorname{dom}(\psi),\ q(h)=\psi(q)\}.
\end{equation}
The policy records each executed world in $K$ and caches its answer. A user response narrows $V_\psi$. For example, specifying completed orders rules out every interpretation that includes pending orders, whatever the missing amount.

\paragraph{Answer Certificate.}
The policy can return $a$ when $V_\psi$ is nonempty and $A(h)=a$ for every $h\in V_\psi$.
Our black-box implementation checks that each surviving world belongs to $K$ and has the same answer.
If $h^\star\in\mathcal H$ and responses are truthful, then $h^\star$ survives every response and the returned answer equals $A(h^\star)$.

\subsection{Answer Signatures}
An interpretation's \emph{answer signature} lists its outputs across repairs. For queries admissible on the same repairs $\mathcal R$, the equivalence relation is
\begin{equation}
 i\equiv_{\mathcal R}j
 \iff \forall r\in\mathcal R,\ A(i,r)=A(j,r).
 \label{eq:stable}
\end{equation}
For order \texttt{O-103} in Figure~\ref{fig:policy}, the two repairs fill in \$0 and \$50, in that order.
Summing all orders gives the signature $(100,150)$ and summing completed orders gives $(100,100)$.
Comparing only the first entries would merge the queries and lose the distinction at \$50.
We use signatures to evaluate candidate pruning.
Grouping worlds to select questions must also preserve response partitions and their probability masses.

\subsection{Question Relevance}
For coordinate questions on a Cartesian product, we call $q$ \emph{locally relevant} if there are surviving worlds $u,v\in V_\psi$ such that
\begin{equation}
 u_j=v_j\ (j\ne q),\qquad A(u)\ne A(v).
 \label{eq:witness}
\end{equation}
The two worlds agree on every other choice, so their different answers establish the effect of changing $q$ alone.
In Figure~\ref{fig:overview}, holding the pending amount at 50 gives such a pair for the intent question.
Comparing all orders at 50 with completed orders at 0 would change both choices and leave the cause of the disagreement unresolved.
Consider an answer given by the XOR of two fair bits.
Conditioning on either bit leaves the answer evenly split, just as conditioning on an unused bit would.
Yet flipping either input with the other held fixed reverses the answer.
The neighboring worlds in Eq.~\ref{eq:witness} expose this dependence.

\method{} requires an observed pair, or \emph{relevance witness}, before asking a question. We count distinct worlds executed ($E$) and questions asked ($Q$). Their weighted cost is $E+cQ$, where $c$ is the cost of a question in execution units. SQL answers are deterministic, hidden worlds equiprobable, and responses to binary questions truthful.

\section{Question Selection}
\label{sec:certificates}
\subsection{Clarification Policy}
The four revenue worlds in Figure~\ref{fig:policy} illustrate how the policy uses evidence from different rounds.
Suppose the all-orders query has returned 100 with the pending amount set to zero and 150 with it set to 50.
These two executions establish a witness for the amount question. If the user supplies 50, only two worlds survive. The policy already knows the all-orders result on that branch and needs one more execution to learn the completed-orders result.
Appendix~\ref{app:examples} traces both responses.

In the third panel, agreement on a known answer across all surviving worlds ($C_a$) lets the assistant return immediately.
A relevance witness ($C_r$) lets the \emph{immediate} policy ask the corresponding question.
The \emph{certified} policy also requires an action certificate ($C_o$) showing that the question maximizes the current EC2 gain.
If the policy can neither answer nor ask, it executes a surviving world whose answer is still unknown.
The fourth panel updates $K$ after execution or filters $V_\psi$ after a user response.
Initially, $V_\psi=\mathcal H$ and $K=\varnothing$.

T1 ranks questions within the interaction. T2 uses Eq.~\ref{eq:stable} to group candidate queries by their answers on the original and modified databases.

EC2 assigns an edge to each unordered pair of worlds with different answers. Its gain is the expected weight of the edges eliminated by a response \citep{golovin2010ec2}.

For $n$ surviving worlds, let $x_{uv}=\mathbf1[A(u)\ne A(v)]$ and let $n_z$ count worlds giving response $z$ to question $q$. With uniform weights,
\begin{align}
 G(q)&=n^{-3}\sum_{\{u,v\}}c_q(u,v)x_{uv},\label{eq:gain}\\
 c_q(u,v)&=\begin{cases}
 n,&q(u)\ne q(v),\\
 n-n_{q(u)},&q(u)=q(v).
 \end{cases}\nonumber
\end{align}
If $u$ and $v$ give different responses, either response eliminates their edge.
If they give the same response $z$, their edge disappears only when the response differs from $z$.

\subsection{Four-Cell Gain Bound}
Assume Boolean answers, a uniform binary Cartesian cube, and equal question costs.
Fixing coordinates through clarification leaves a smaller cube.
For two unanswered coordinates $q,r$, group the surviving worlds by their responses $(q,r)$.
Let $a,b,c,d$ count the positive answers in cells $00,01,10,11$, respectively.

\paragraph{Proposition 1.}
The scaled gain difference equals
\begin{equation}
 n^2\bigl(G(q)-G(r)\bigr)=(a-d)(b-c).
 \label{eq:fourcell}
\end{equation}
Each question bisects the cube. The positive counts in the two halves are $(a+b,c+d)$ for $q$ and $(a+c,b+d)$ for $r$.
Both remove every edge that crosses their split and, in expectation, half the edges within the two halves.
Since the total edge weight is fixed, the question that leaves fewer edges within its halves has greater gain.
Comparing those within-half counts gives the product in Eq.~\ref{eq:fourcell} (Appendix~\ref{app:proof}).

Each execution narrows the range of one cell count.
For cell $00$, the lower bound $a_-$ is its number of observed positives and the upper bound $a_+$ adds its number of unknown answers.
Applying the same bounds to each cell gives
\begin{align}
 a-d&\in[a_- - d_+,a_+ - d_-],\\
 b-c&\in[b_- - c_+,b_+ - c_-].
\end{align}
The factors depend on disjoint pairs of cells.
Assigning the unknown Boolean answers within each cell therefore attains every combination of interval endpoints.
The smallest and largest of the four endpoint products give exact bounds on the scaled gain difference.
If the lower bound for $q$ against every competitor is nonnegative, $q$ maximizes the gain under every completion.
The policy asks the first certified maximizer with a relevance witness.

\subsection{Execution Order}
We compare the four-cell bound with a \emph{partition bound} that groups executed worlds by answer. An unknown world can join one of these groups or form a new one.
For each unknown world, the relaxation chooses the class assignment that gives the smallest contribution to the gain difference.
These assignments can be incompatible when two unknown worlds share an edge.
The four-cell bound counts each Boolean label in a single response cell.
Appendix~\ref{app:proof} derives the relaxation.

When a policy needs another execution, \emph{fixed} order takes the first unknown surviving world.
\emph{Witness} order prefers worlds with the most distinct labels among known neighbors one coordinate away. It breaks ties by neighbor count and then by fixed order.
If the neighbors have both Boolean labels, either outcome supplies a witness.

\paragraph{Stopping Conditions.}
Each execution reveals a new label and each question fixes a coordinate.
On a fully known nonconstant cube, a path between worlds with different answers must contain an edge whose endpoints have different labels.
The coordinate on that edge is relevant.
An irrelevant coordinate splits the positive labels evenly and has minimum EC2 gain.
Thus some relevant coordinate attains the maximum, even if all gains tie.
Either policy can then ask that question.

On a constant map, every question is irrelevant. The policy must execute all worlds because any unseen answer could differ.

\section{Experimental Evaluation}
\label{sec:experiments}
\subsection{Tasks And Policies}
We test question selection and candidate pruning separately (Table~\ref{tab:evidence}). Sales Dev uses the original SQL outputs. Sales Test measures policy costs on Boolean versions of the same three queries with new seeds.
AmbiQT Swap tests whether two reference queries still agree after a value swap. AmbiQT Candidates measures which reference answers survive pruning in a sample from that audit.
Boolean Stress isolates six patterns of answer dependence, and Boolean Cube enumerates partially labeled maps.
Score Sums times the aggregation of worlds with equal integer scores (Appendix~\ref{app:diagnostics}).

\begin{table}[htbp]
\centering\small
\begin{tabular}{@{}lr@{}}
\toprule
Evaluation set & Size\\
\midrule
Sales Dev & 300 tasks\\
Sales Test & 120 tasks\\
Boolean Stress & 48 fixtures\\
Boolean Cube & $3^8$ maps\\
AmbiQT Swap & 1,240 items\\
AmbiQT Candidates & 155 questions, 2 models\\
Score Sums & 3 seeds per size\\
\bottomrule
\end{tabular}
\caption{Evaluation sets and sample sizes.}
\label{tab:evidence}
\end{table}
\ifdefined\begin{table*}[t]
\centering\small
\setlength{\tabcolsep}{3.6pt}
\begin{tabular}{llrrrrrrr}
\toprule
 & & Relevant & \multicolumn{2}{c}{Exact bound} & \multicolumn{2}{c}{Partition bound} & \multicolumn{2}{c}{Immediate witness} \\
\cmidrule(lr){4-5}\cmidrule(lr){6-7}\cmidrule(lr){8-9}
Task family & Metric & EC2 & Fixed & Witness & Fixed & Witness & Fixed & Witness \\
\midrule
Leading region ($n=40$) & $E\downarrow$ & 16.00 & 15.16 & 15.16 & 15.68 & 15.67 & \textbf{13.83} & \textbf{13.83} \\
 & $Q$ & 0.52 & 0.52 & 0.52 & 0.52 & 0.52 & 0.57 & 0.57 \\
 & $C_{10}\downarrow$ & 21.19 & 20.35 & 20.35 & 20.87 & 20.86 & \textbf{19.52} & \textbf{19.52} \\
\midrule
Customer count ($n=40$) & $E\downarrow$ & 16.00 & \textbf{12.50} & \textbf{12.50} & 14.00 & 14.00 & \textbf{12.50} & \textbf{12.50} \\
 & $Q$ & 1.00 & 1.00 & 1.00 & 1.00 & 1.00 & 1.00 & 1.00 \\
 & $C_{10}\downarrow$ & 26.00 & \textbf{22.50} & \textbf{22.50} & 24.00 & 24.00 & \textbf{22.50} & \textbf{22.50} \\
\midrule
Order threshold ($n=40$) & $E\downarrow$ & 16.00 & 13.75 & 13.78 & 14.95 & 15.03 & \textbf{13.22} & \textbf{13.22} \\
 & $Q$ & 0.91 & 0.91 & 0.91 & 0.91 & 0.91 & 0.94 & 0.94 \\
 & $C_{10}\downarrow$ & 25.12 & 22.87 & 22.90 & 24.07 & 24.16 & \textbf{22.59} & \textbf{22.59} \\
\midrule
All tasks ($n=120$) & $E\downarrow$ & 16.00 & 13.80 & 13.81 & 14.88 & 14.90 & \textbf{13.18} & \textbf{13.18} \\
 & $Q$ & 0.81 & 0.81 & 0.81 & 0.81 & 0.81 & 0.84 & 0.84 \\
 & $C_{10}\downarrow$ & 24.10 & 21.91 & 21.92 & 22.98 & 23.00 & \textbf{21.54} & \textbf{21.54} \\
\bottomrule
\end{tabular}
\caption{Sales Test costs. Bold marks the lowest $E$ and $C_{10}=E+10Q$ within each group.}
\label{tab:main-expanded}
\end{table*}
\begin{table*}[t]
\centering\small
\setlength{\tabcolsep}{3.6pt}
\begin{tabular}{llrrrrrrr}
\toprule
 & & Relevant & \multicolumn{2}{c}{Exact bound} & \multicolumn{2}{c}{Partition bound} & \multicolumn{2}{c}{Immediate witness} \\
\cmidrule(lr){4-5}\cmidrule(lr){6-7}\cmidrule(lr){8-9}
Task family & Metric & EC2 & Fixed & Witness & Fixed & Witness & Fixed & Witness \\
\midrule
Leading region ($n=40$) & $E\downarrow$ & 16.00 & 15.16 & 15.16 & 15.68 & 15.67 & \textbf{13.83} & \textbf{13.83} \\
 & $Q$ & 0.52 & 0.52 & 0.52 & 0.52 & 0.52 & 0.57 & 0.57 \\
 & $C_{10}\downarrow$ & 21.19 & 20.35 & 20.35 & 20.87 & 20.86 & \textbf{19.52} & \textbf{19.52} \\
\midrule
Customer count ($n=40$) & $E\downarrow$ & 16.00 & \textbf{12.50} & \textbf{12.50} & 14.00 & 14.00 & \textbf{12.50} & \textbf{12.50} \\
 & $Q$ & 1.00 & 1.00 & 1.00 & 1.00 & 1.00 & 1.00 & 1.00 \\
 & $C_{10}\downarrow$ & 26.00 & \textbf{22.50} & \textbf{22.50} & 24.00 & 24.00 & \textbf{22.50} & \textbf{22.50} \\
\midrule
Order threshold ($n=40$) & $E\downarrow$ & 16.00 & 13.75 & 13.78 & 14.95 & 15.03 & \textbf{13.22} & \textbf{13.22} \\
 & $Q$ & 0.91 & 0.91 & 0.91 & 0.91 & 0.91 & 0.94 & 0.94 \\
 & $C_{10}\downarrow$ & 25.12 & 22.87 & 22.90 & 24.07 & 24.16 & \textbf{22.59} & \textbf{22.59} \\
\midrule
All tasks ($n=120$) & $E\downarrow$ & 16.00 & 13.80 & 13.81 & 14.88 & 14.90 & \textbf{13.18} & \textbf{13.18} \\
 & $Q$ & 0.81 & 0.81 & 0.81 & 0.81 & 0.81 & 0.84 & 0.84 \\
 & $C_{10}\downarrow$ & 24.10 & 21.91 & 21.92 & 22.98 & 23.00 & \textbf{21.54} & \textbf{21.54} \\
\bottomrule
\end{tabular}
\caption{Sales Test costs. Bold marks the lowest $E$ and $C_{10}=E+10Q$ within each group.}
\label{tab:main-expanded}
\end{table*}
\fi

For each of 40 Sales Test seeds, we generate a leading-region query, a customer-count query, and an order-count query.
One intent choice and three decisions about duplicate records define 16 worlds for each query.
To test the Boolean bound, we ask whether East is among the leading regions, whether the customer count is at least six, and whether the thresholded order count is at least five.
We fixed the seeds and Boolean predicates before evaluation.
Each world serves once as the simulator's hidden world, giving 1,920 episodes per policy. We precompute the answers in SQLite and reveal each to the policy only when it requests that execution.
We compute paired confidence intervals using 2,000 bootstrap samples of the 40 seeds, keeping the three queries from each seed together.
Appendix~\ref{app:selector} reports the original SQL results.

\paragraph{Policy Configurations.}
We evaluate ten configurations.
The anchor policy returns the answer from one execution.
Ask-all resolves every coordinate and then executes the remaining world.
Full EC2 executes all worlds before selecting questions.
Full relevant EC2 also checks each question for a local witness (Eq.~\ref{eq:witness}).
The other six configurations pair exact, partition, or immediate selection with fixed or witness execution order.
Table~\ref{tab:main-expanded} compares the seven policies that require a witness for every question and agreement before returning an answer.
Table~\ref{tab:contract-audit} reports violations for all ten.

\ifdefined\begin{table*}[t]
\centering\small
\setlength{\tabcolsep}{3.6pt}
\begin{tabular}{llrrrrrrr}
\toprule
 & & Relevant & \multicolumn{2}{c}{Exact bound} & \multicolumn{2}{c}{Partition bound} & \multicolumn{2}{c}{Immediate witness} \\
\cmidrule(lr){4-5}\cmidrule(lr){6-7}\cmidrule(lr){8-9}
Task family & Metric & EC2 & Fixed & Witness & Fixed & Witness & Fixed & Witness \\
\midrule
Leading region ($n=40$) & $E\downarrow$ & 16.00 & 15.16 & 15.16 & 15.68 & 15.67 & \textbf{13.83} & \textbf{13.83} \\
 & $Q$ & 0.52 & 0.52 & 0.52 & 0.52 & 0.52 & 0.57 & 0.57 \\
 & $C_{10}\downarrow$ & 21.19 & 20.35 & 20.35 & 20.87 & 20.86 & \textbf{19.52} & \textbf{19.52} \\
\midrule
Customer count ($n=40$) & $E\downarrow$ & 16.00 & \textbf{12.50} & \textbf{12.50} & 14.00 & 14.00 & \textbf{12.50} & \textbf{12.50} \\
 & $Q$ & 1.00 & 1.00 & 1.00 & 1.00 & 1.00 & 1.00 & 1.00 \\
 & $C_{10}\downarrow$ & 26.00 & \textbf{22.50} & \textbf{22.50} & 24.00 & 24.00 & \textbf{22.50} & \textbf{22.50} \\
\midrule
Order threshold ($n=40$) & $E\downarrow$ & 16.00 & 13.75 & 13.78 & 14.95 & 15.03 & \textbf{13.22} & \textbf{13.22} \\
 & $Q$ & 0.91 & 0.91 & 0.91 & 0.91 & 0.91 & 0.94 & 0.94 \\
 & $C_{10}\downarrow$ & 25.12 & 22.87 & 22.90 & 24.07 & 24.16 & \textbf{22.59} & \textbf{22.59} \\
\midrule
All tasks ($n=120$) & $E\downarrow$ & 16.00 & 13.80 & 13.81 & 14.88 & 14.90 & \textbf{13.18} & \textbf{13.18} \\
 & $Q$ & 0.81 & 0.81 & 0.81 & 0.81 & 0.81 & 0.84 & 0.84 \\
 & $C_{10}\downarrow$ & 24.10 & 21.91 & 21.92 & 22.98 & 23.00 & \textbf{21.54} & \textbf{21.54} \\
\bottomrule
\end{tabular}
\caption{Sales Test costs. Bold marks the lowest $E$ and $C_{10}=E+10Q$ within each group.}
\label{tab:main-expanded}
\end{table*}
\else\begin{table*}[t]
\centering\small
\setlength{\tabcolsep}{3.6pt}
\begin{tabular}{llrrrrrrr}
\toprule
 & & Relevant & \multicolumn{2}{c}{Exact bound} & \multicolumn{2}{c}{Partition bound} & \multicolumn{2}{c}{Immediate witness} \\
\cmidrule(lr){4-5}\cmidrule(lr){6-7}\cmidrule(lr){8-9}
Task family & Metric & EC2 & Fixed & Witness & Fixed & Witness & Fixed & Witness \\
\midrule
Leading region ($n=40$) & $E\downarrow$ & 16.00 & 15.16 & 15.16 & 15.68 & 15.67 & \textbf{13.83} & \textbf{13.83} \\
 & $Q$ & 0.52 & 0.52 & 0.52 & 0.52 & 0.52 & 0.57 & 0.57 \\
 & $C_{10}\downarrow$ & 21.19 & 20.35 & 20.35 & 20.87 & 20.86 & \textbf{19.52} & \textbf{19.52} \\
\midrule
Customer count ($n=40$) & $E\downarrow$ & 16.00 & \textbf{12.50} & \textbf{12.50} & 14.00 & 14.00 & \textbf{12.50} & \textbf{12.50} \\
 & $Q$ & 1.00 & 1.00 & 1.00 & 1.00 & 1.00 & 1.00 & 1.00 \\
 & $C_{10}\downarrow$ & 26.00 & \textbf{22.50} & \textbf{22.50} & 24.00 & 24.00 & \textbf{22.50} & \textbf{22.50} \\
\midrule
Order threshold ($n=40$) & $E\downarrow$ & 16.00 & 13.75 & 13.78 & 14.95 & 15.03 & \textbf{13.22} & \textbf{13.22} \\
 & $Q$ & 0.91 & 0.91 & 0.91 & 0.91 & 0.91 & 0.94 & 0.94 \\
 & $C_{10}\downarrow$ & 25.12 & 22.87 & 22.90 & 24.07 & 24.16 & \textbf{22.59} & \textbf{22.59} \\
\midrule
All tasks ($n=120$) & $E\downarrow$ & 16.00 & 13.80 & 13.81 & 14.88 & 14.90 & \textbf{13.18} & \textbf{13.18} \\
 & $Q$ & 0.81 & 0.81 & 0.81 & 0.81 & 0.81 & 0.84 & 0.84 \\
 & $C_{10}\downarrow$ & 24.10 & 21.91 & 21.92 & 22.98 & 23.00 & \textbf{21.54} & \textbf{21.54} \\
\bottomrule
\end{tabular}
\caption{Sales Test costs. Bold marks the lowest $E$ and $C_{10}=E+10Q$ within each group.}
\label{tab:main-expanded}
\end{table*}
\fi

\subsection{Policy Costs}
\begin{figure*}[t]
\centering
\includegraphics[width=0.92\textwidth]{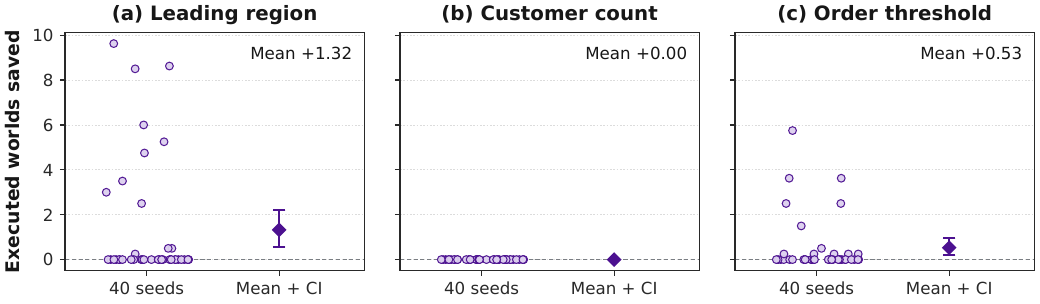}
\par\vspace{-3pt}
\caption{Executions saved by immediate selection with witness order relative to exact selection with fixed order. Diamonds show means and bars give 95\% intervals from resampling seeds.}
\label{fig:paired}
\vspace{-3pt}
\end{figure*}

With fixed execution order, the exact bound uses 13.80 executions per episode, compared with 14.88 for the partition bound and 16 for full relevant EC2 (Table~\ref{tab:main-expanded}).
Among the seven policies in the table, immediate selection uses the fewest executions, 13.18 per episode.
Waiting to certify an EC2 maximizer costs an additional 0.619 executions on average (95\% paired interval $[0.346,0.942]$) and reduces mean questions from 0.835 to 0.810.
Immediate selection saves 1.325 executions on leading-region queries and 0.531 on order thresholds relative to exact selection with fixed order (Figure~\ref{fig:paired}, using witness order for immediate selection).
They tie on every customer-count seed at a mean of 12.50.

Changing the execution order has little effect on these means. The immediate policy uses 13.18 executions with either order. Searching the known neighbors for the most informative world increases its decision time from 0.168 to 0.516 milliseconds per episode.
For the exact policy, witness order costs 13.81 executions, compared with 13.80 under fixed order.
These timings cover the policy and its lookup of cached answers. Query execution on a live database would add a separate cost to each requested world.

The seven policies give the correct answer and ask only witnessed questions on all 1,920 sales episodes per policy.
The anchor policy gives 580 wrong answers.
Ask-all answers correctly after four questions, of which 79.11\% are irrelevant.
It asks every question before its first execution, so none has an observed witness when asked.

\begin{figure}[htbp]
\centering
\includegraphics[width=\columnwidth]{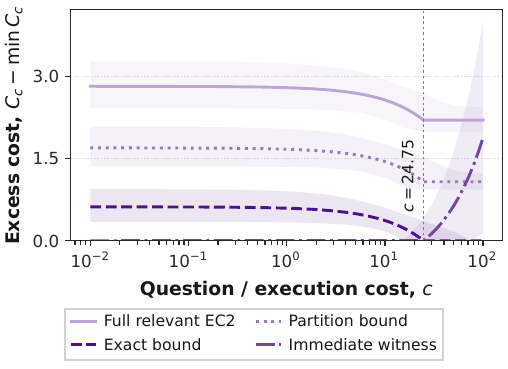}
\par\vspace{-3pt}
\caption{Excess cost at each question weight. Shading gives pointwise 95\% paired bootstrap intervals.}
\label{fig:sensitivity}
\vspace{-3pt}
\end{figure}

Figure~\ref{fig:sensitivity} shows how policy costs vary with the question weight.
With fixed order the mean cost difference between exact and immediate selection is $0.61875-0.025c$.
At $c=10$, immediate selection costs 21.54 and exact selection costs 21.91.
Exact selection costs less above $c=24.75$.
Appendix~\ref{app:diagnostics} reports timings.

\subsection{Certificate Quality}
\begin{figure*}[t]
\centering
\includegraphics[width=0.92\textwidth]{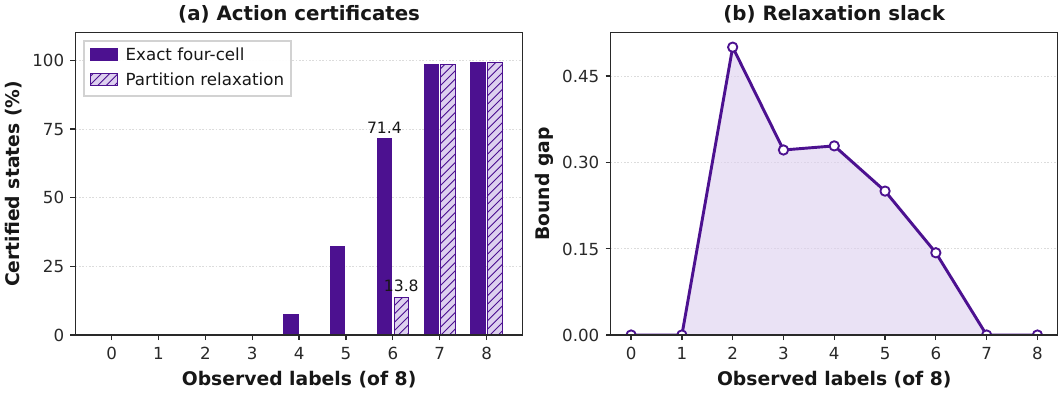}
\par\vspace{-2pt}
\caption{Witnessed EC2 maximizers (left) and mean bound gap in edge-cut units (right).}
\label{fig:bounds}
\vspace{-1pt}
\end{figure*}
To measure what the bound itself contributes, we enumerate all $3^8=6{,}561$ partial answer maps on three binary coordinates.
Each of the eight worlds can have a known positive answer, a known negative answer, or an unknown answer.
We check Eq.~\ref{eq:fourcell} against direct edge counts and apply both bounds to each partial map.
With six answers known, the exact bound certifies a maximizer with a relevance witness in 71.4\% of maps, compared with 13.8\% for the partition bound (Figure~\ref{fig:bounds}).

Our six stress tests cover a constant, a single relevant bit, XOR with nuisance coordinates, a rare conjunction, a weighted threshold, and branching on intent.
Eight seeds give 48 fixtures, some with repeated answer maps (Figure~\ref{fig:stress}).
On the XOR fixture, full EC2 breaks ties by fixed order and spends half its questions on irrelevant coordinates.
Requiring a relevance witness removes these questions.
On a constant map, every policy that requires a witness executes all 16 worlds and asks zero questions.
The cost changes when the query has known structure.
In Score Sums, worlds with the same integer score share a count.
With 256 binary repairs and one binary intent, the score histograms for $2^{257}$ worlds take 0.55 to 0.59 seconds to compute across three seeds.
Appendix~\ref{app:diagnostics} compares this computation with enumeration.

\subsection{Sensitivity To Repairs}
We examine all 1,240 validation items in the column-synonym category of AmbiQT \citep{bhaskar-etal-2023-benchmarking}, using the associated Spider databases \citep{yu2018spider}.
Spider-Syn changes schema terms in the question \citep{gan-etal-2021-towards}, and Dr.Spider perturbs databases, questions, and SQL \citep{chang2023drspider}. We hold both reference queries fixed and swap two values in the second synonym column.
We choose the rows deterministically from distinct non-null values before examining query outputs.
Twenty items fail to execute and 14 more lack a pair of eligible values, leaving 1,206 cases.
We execute both queries on both states for each eligible case.

The two reference queries agree on the observed database and disagree after the swap in 424 cases (35.16\% of eligible interventions).
In the other 782 cases, all four worlds give the same answer.
Swapping two values preserves the column's multiset, so a whole-column aggregate can remain constant even as the values associated with particular rows change.
In each of the 424 affected cases, pruning on the observed database gives the right answer in three of the four worlds.
It makes 424 errors in $4\times1{,}206$ equally weighted trials, yielding 91.21\% accuracy.
Given both reference queries and the complete answer map, EC2 averages 1.5 questions on the 424 affected cases. On the other 782 it can return the common answer immediately. It answers correctly in every world, averaging 0.5274 questions.
Figure~\ref{fig:audit} gives the database counts.

\subsection{Reference Signature Recovery}
We sample up to eight distinct questions per database by a deterministic hash, giving 155 questions across 20 databases.
Doubao Lite and GPT Sol each receive the question and schema and propose up to four SQL interpretations in one request.
We discard queries that fail on either state. In generation order, observed-state pruning keeps the first query for each distinct original answer. Two-state pruning keeps the first for each distinct pair of original and modified answers.

\begin{table*}[t]
\centering\small
\setlength{\tabcolsep}{3.5pt}
\begin{tabular}{llrrrrrrrrr}
\toprule
 & & \multicolumn{3}{c}{All cases ($n=155$)} & \multicolumn{3}{c}{Two signatures ($n=53$)} & \multicolumn{3}{c}{One signature ($n=102$)} \\
\cmidrule(lr){3-5}\cmidrule(lr){6-8}\cmidrule(lr){9-11}
Model & Pruning & Rec. & Prec. & Full & Rec. & Prec. & Full & Rec. & Prec. & Full \\
\midrule
Doubao Lite & Observed & 59.03 & 47.37 & 44.52 & 42.45 & 51.73 & 0.00 & 67.65 & 45.10 & 67.65 \\
 & Two-state & 67.74 & 48.01 & 61.94 & 67.92 & 55.82 & 50.94 & 67.65 & 43.95 & 67.65 \\
\midrule
GPT Sol & Observed & 59.03 & 59.89 & 43.87 & 44.34 & 73.43 & 0.00 & 66.67 & 52.86 & 66.67 \\
 & Two-state & 69.03 & 59.46 & 63.87 & 73.58 & 74.69 & 58.49 & 66.67 & 51.55 & 66.67 \\
\bottomrule
\end{tabular}
\caption{Reference recovery (\%) on AmbiQT Candidates.}
\label{tab:model-expanded}
\end{table*}

For each question, recall is the fraction of reference signatures recovered, and precision is the fraction of retained signatures that match a reference. Full recovery requires every reference signature. Table~\ref{tab:model-expanded} averages each measure over questions.
Both rules retain the first query for each original answer. The recall difference measures the coverage lost by pruning before the swap.

Two-state pruning raises mean recall by 8.71 percentage points for Doubao Lite (95\% paired database bootstrap interval $[4.69,12.81]$) and 10.00 for GPT Sol ($[5.94,14.14]$).
All gains occur on the 53 questions with two reference signatures. Since both yield the same original answer, observed-state pruning retains at most one. After the swap, both signatures survive on 27 of these questions for Doubao Lite and 31 for GPT Sol.
Mean recall increases in 11 of 20 databases for Doubao Lite and in 12 for GPT Sol (Figure~\ref{fig:database}).

Two-state pruning adds 39 signatures for Doubao Lite and 45 for GPT Sol. Of these, 12 and 14 match neither reference. Mean precision changes by $+0.65$ and $-0.43$ points, respectively. Both paired 95\% intervals include zero.
Of the 208 reference signatures across questions, 67 for Doubao Lite and 62 for GPT Sol are already absent before pruning.
For songs by singers older than average, GPT Sol proposed both \texttt{track\_name} and \texttt{composition\_name}. Swapping two composition names separated the lists (Appendix~\ref{app:song}).

\subsection{SQL Error Analysis}
\label{sec:sql-errors}
Before pruning, 276 Doubao Lite queries and 177 GPT Sol queries executed on both states but matched neither reference signature. A further 11 and six, respectively, failed to execute on at least one state.
We sampled 50 executable mismatches per model by hash. The sample spans 73 questions and 20 databases. With model names hidden, OpenAI Codex compared each query with both references in one pass and labeled its discrepancies from the closer one (Table~\ref{tab:sql-errors}). Since the comparison rule affects measured accuracy \citep{renggli2025evaluation}, we also tested the effect of column order.

\begin{table}[htbp]
\centering\small
\begin{tabular}{@{}lrrr@{}}
\toprule
Discrepancy & Doubao Lite & GPT Sol & Pooled \\
\midrule
Output columns & 17 & 29 & 46 \\
Source or join & 15 & 13 & 28 \\
Aggregation & 13 & 14 & 27 \\
Predicate & 11 & 8 & 19 \\
Ranking or ties & 5 & 9 & 14 \\
\bottomrule
\end{tabular}
\caption{Counts among 50 sampled SQL mismatches per model. Of the 100 queries, 32 had multiple labels.}
\label{tab:sql-errors}
\end{table}

Output columns differed in 46 of the 100 queries. For example, a query returned opening year and staff size in the order the question requested, while the reference returned staff size first. Across all 453 executable mismatches, we tested whether a fixed permutation of the columns reproduced either reference on both states. This recovered a match for 17 of 276 Doubao Lite queries and 15 of 177 GPT Sol queries. The choice of join could change who appeared in an answer. A left join counted friends for all 16 students, including two with no friendship record. Both references used an inner join and reported counts for the other 14 students.

\section{Main Conclusions}
\method{} asks about a choice only when two SQL executions show that it can change the answer. On 120 Boolean sales tasks, the certified policy averaged 13.80 executions per episode with the four-cell bound and 14.88 with the partition bound. Asking at the first witness saved another 0.62 executions at the cost of 0.025 more questions per episode. On AmbiQT, two-state pruning preserved a second reference signature on 27 of 155 questions for Doubao Lite and 31 for GPT Sol. In each case the queries agreed on the original records and gave different answers after the swap.

\section*{Limitations}
Answer correctness assumes that the supplied candidates include the intended query and the true database state.
The AmbiQT audit uses one value swap in one category.
Estimating their frequency requires data on actual repairs.

The sales tasks use new seeds from the development generator.
The simulator answers truthfully and assigns the same cost to each question.
Measuring the effort and reliability of real responses requires a user study.
The exact bound assumes Boolean answers on a uniform binary cube.
Dependencies among repairs or unequal probabilities change the gain calculation.
\section*{Ethical Considerations}
The study uses public databases, generated sales records, and simulated users. Models generate candidates from schemas and questions alone.
\section*{Acknowledgements}
We used OpenAI Codex to organize and revise the manuscript, draft the explanations in Appendix~\ref{app:examples}, and edit the \LaTeX{} source.
We consulted Doubao Lite and GPT Sol on question selection and the EC2 derivation. We checked the gain identity by exhaustive enumeration (Section~\ref{sec:experiments}). Both models also generated candidate SQL queries for the AmbiQT evaluation (Appendix~\ref{app:repro}).
\bibliography{references}
\appendix
\section{Certificate Proofs}
\label{app:proof}
\subsection{Four-Cell Identity}
Let $n=2m$ surviving worlds have $P$ positive Boolean answers.
There are $T=P(n-P)$ edges between opposite labels.
For a coordinate $q$, let $p_0,p_1$ be the positive counts in its two halves.
The number of edges within the halves is
\begin{equation}
 W_q=p_0(m-p_0)+p_1(m-p_1).
\end{equation}
A cross-half edge is cut under either response. A within-half edge is cut with probability $1/2$.
Each edge has weight $1/n^2$, giving $n^2G(q)=T-W_q/2$.
Since $p_0+p_1=P$, the only term in the gain that depends on the split is $(p_0^2+p_1^2)/2$.
For $q$ the two counts are $a+b,c+d$, and for $r$ they are $a+c,b+d$.
Subtracting the gains gives
\begin{align*}
 n^2(G(q)-G(r))
 &=\tfrac12\big[(a+b)^2+(c+d)^2\\
 &\qquad -(a+c)^2-(b+d)^2\big]\\
 &=(a-d)(b-c).
\end{align*}

\paragraph{Attainable Bounds.}
The observed positives in a cell give its smallest possible final count. Assigning every unknown label in that cell to one gives its largest count.
Both extremes of $a-d$ are therefore attainable, independently of either extreme of $b-c$.
The minimum and maximum of their product occur at corners of this rectangle, and every corner is realized by a completion.
A question with a nonnegative lower bound against every competitor maximizes gain under every completion.

\subsection{Partition Relaxation}
The \emph{partition bound} accommodates any number of answer classes.
Let $d_{uv}=c_q(u,v)-c_r(u,v)$.
A pair of known answers contributes $d_{uv}x_{uv}$ to the unnormalized gain difference.
A pair of unknown answers contributes at least $\min(0,d_{uv})$.
For an unknown world $u$, group its known neighbors by their answers $s$.
Write $g_{u,s}=\sum_{v:A(v)=s}d_{uv}$ for the sum over known neighbors in class $s$, and $T_u=\sum_sg_{u,s}$.
The contribution from all known neighbors of $u$ is at least
\begin{equation}
 \ell_u=\min\bigl(\{T_u\}\cup\{T_u-g_{u,s}:s\text{ known}\}\bigr).
\end{equation}
If $u$ has a new answer, every known class contributes.
If it belongs to a known class $s$, only the edges to that class drop out.
The bound enforces this choice for each unknown world separately.
The individual minima can correspond to incompatible assignments across worlds, which makes the partition bound a relaxation.

\subsection{Policy Termination}
Each execution reveals a new answer and each question fixes a new coordinate.
On a fully executed nonconstant cube, a path between two different answers must contain neighbors with different answers, so at least one coordinate is locally relevant.
An irrelevant coordinate has identical positive counts in its two halves, which minimizes EC2 gain.
If its gain is maximal, all gains tie and a relevant coordinate also attains the maximum.
Thus both policies can ask a witnessed question on a fully known nonconstant cube.
They terminate by returning the common answer once the answer certificate holds.
To see why an irrelevant coordinate has minimum gain, observe that each positive world on one side has a positive partner on the other.
This gives $p_0=p_1=P/2$, which minimizes $p_0^2+p_1^2$ for fixed $P$.
On a constant map the policy has no admissible question. Until every world is known, changing one unseen label remains consistent with the observations.
The worst case therefore requires $|\mathcal H|$ executions.

\section{Complete SQL Answers}
\label{app:selector}
The 300 development tasks use complete SQL results, including counts and sets of leading regions.
Table~\ref{tab:e001} compares selectors with access to the full answer map.
Each simulated user responds according to one of the supplied worlds.
We measure how often a selector returns the correct result and how many questions it asks.

\begin{table}[t]
\centering
\small
\begin{tabular}{lrr}
\toprule
Strategy & Correct (\%) & Questions \\
\midrule
No clarification & 38.46 & 0.0000 \\
Ask all & 100.00 & 4.0000 \\
Fixed order & 100.00 & 2.1850 \\
Random order & 100.00 & 2.4900 \\
World entropy & 100.00 & 2.1850 \\
Answer entropy & 100.00 & 1.8079 \\
EC2 & 100.00 & 1.8079 \\
Local check & 98.98 & 1.7058 \\
Observed then repair & 98.88 & 2.0246 \\
Repair first & 100.00 & 2.9354 \\
Exact optimal & 100.00 & 1.8067 \\
\bottomrule
\end{tabular}
\caption{Question selection on Sales Dev.}
\label{tab:e001}
\end{table}

The anchor policy answers immediately, and ask-all resolves every coordinate.
Fixed and random orders stop as soon as all surviving answers agree.
World entropy and answer entropy choose the coordinate with the largest expected reduction in their respective uncertainties.
EC2 chooses the largest expected reduction in edge weight between different answer classes.
A local check probes changes to one coordinate at the anchor.
Observed-then-repair merges queries on the observed data. Repair-first asks about all repairs first.

An exact decision tree gives the minimum mean question count of 1.8067.
Answer entropy and EC2 both average 1.8079 questions with correct answers throughout.
The local check asks fewer questions (1.7058) and gives a correct answer in 98.98\% of episodes.
Resolving all repairs before intent costs 2.9354 questions on average.

\section{Policy Diagnostics}
\label{app:diagnostics}

\begin{table*}[!t]
\centering\small
\setlength{\tabcolsep}{3.7pt}
\begin{tabular}{lrrrrrrrr}
\toprule
 & \multicolumn{4}{c}{Sales Test} & \multicolumn{4}{c}{Boolean Stress} \\
\cmidrule(lr){2-5}\cmidrule(lr){6-9}
Configuration & Wrong & Irr.Q & Unpr.Q & No cert. & Wrong & Irr.Q & Unpr.Q & No cert. \\
\midrule
Anchor answer & 580 & n/a & n/a & 1920 & 268 & n/a & n/a & 768 \\
Ungated ask-all & 0 & 6076 & 7680 & 0 & 0 & 1776 & 3072 & 0 \\
Full EC2 & 0 & 0 & 0 & 0 & 0 & 256 & 256 & 0 \\
Full relevant EC2 & 0 & 0 & 0 & 0 & 0 & 0 & 0 & 0 \\
Exact / fixed & 0 & 0 & 0 & 0 & 0 & 0 & 0 & 0 \\
Exact / witness & 0 & 0 & 0 & 0 & 0 & 0 & 0 & 0 \\
Partition / fixed & 0 & 0 & 0 & 0 & 0 & 0 & 0 & 0 \\
Partition / witness & 0 & 0 & 0 & 0 & 0 & 0 & 0 & 0 \\
Immediate / fixed & 0 & 0 & 0 & 0 & 0 & 0 & 0 & 0 \\
Immediate / witness & 0 & 0 & 0 & 0 & 0 & 0 & 0 & 0 \\
\bottomrule
\end{tabular}
\caption{Answer and question violations by policy.}
\label{tab:contract-audit}
\end{table*}

Table~\ref{tab:contract-audit} counts decisions across 1,920 sales episodes and 768 stress episodes per configuration.
\emph{Wrong} counts incorrect answers and \emph{No cert.} counts returns before the surviving answers were known to agree.
\emph{Irr.Q} counts questions for which the full answer map contains no local witness.
\emph{Unpr.Q} counts questions for which the policy had yet to observe a witness.
The anchor policy asks zero questions.
Full EC2 asks 256 irrelevant questions on Boolean Stress.

\begin{figure*}[tp]
\centering
\includegraphics[width=.95\textwidth]{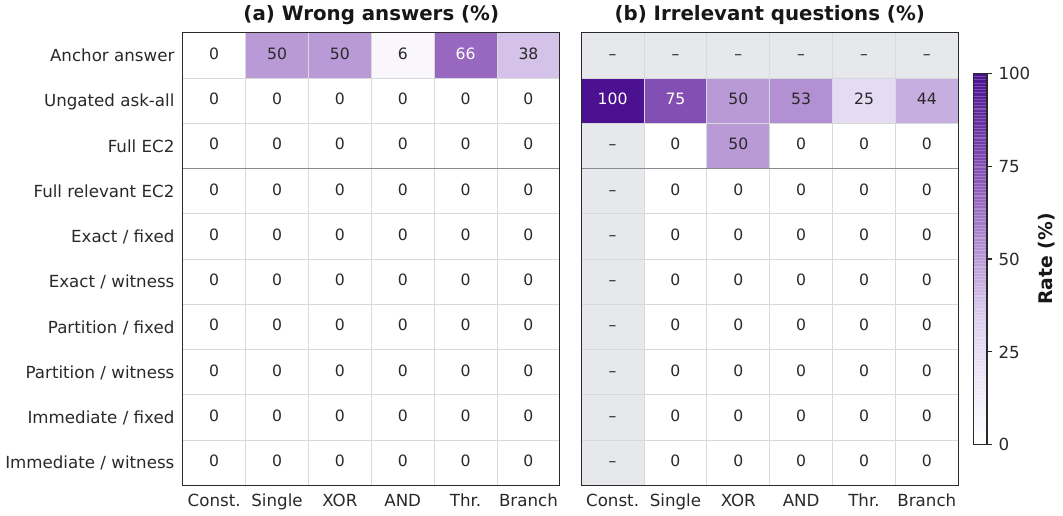}
\caption{Answer and question relevance failures on six Boolean functions. Question rates are undefined for policies that ask no questions.}
\label{fig:stress}
\end{figure*}
Figure~\ref{fig:stress} gives failure rates by Boolean function.
White cells indicate zero violations, and gray cells mark undefined rates.
On the nuisance XOR fixture, a fixed tie order can favor an irrelevant coordinate even after the full answer map is known.
A local witness distinguishes an input to XOR from a nuisance coordinate with the same EC2 gain.

On the development tasks, we also tested two looser criteria.
Bounding each pair of worlds independently required 15.3913 executions per episode to reproduce the full EC2 question sequence.
Grouping known answers by class reduced this to 14.9000 with the same sequence.
A separate heuristic counted unknown worlds and pairs with different answers to guide executions and questions.
It allowed a question whenever two worlds with different answers gave different responses to it.
If the worlds differ on two coordinates, either may explain the answer change.

\subsection{Policy Costs}
\begin{figure*}[t]
\centering
\includegraphics[width=\textwidth]{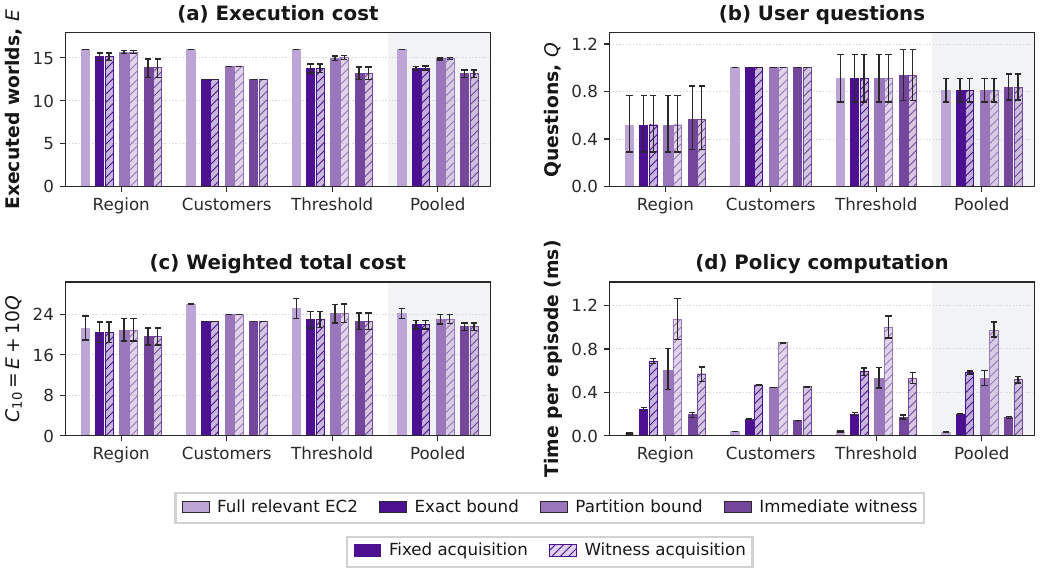}
\caption{Executions, questions, weighted cost, and policy computation by query family. Error bars show 95\% intervals from resampling seeds.}
\label{fig:frontier}
\end{figure*}
Choosing the next execution by its known neighbors adds computation with little change in executions (Figure~\ref{fig:frontier}).
Under the exact bound, switching from fixed to witness order changes the mean from 13.8021 to 13.8135 and increases decision time from 0.200 to 0.582 milliseconds.
With the partition bound, the corresponding means are 14.8771 and 14.9000 executions, with decision times of 0.526 and 0.974 milliseconds.
Both immediate policies use 13.1833 executions.
Searching the neighbors increases their decision time from 0.168 to 0.516 milliseconds.
These timings cover the policy and its lookup of cached answers.

\paragraph{Question Cost.}
In units where one execution costs one and one question costs $c$, the mean cost of exact selection minus that of immediate selection is $0.61875-0.025c$ under fixed order.
At $c=10$, Table~\ref{tab:main-expanded} gives mean costs of 21.91 for exact and 21.54 for immediate selection.
The two are equal at $c=24.75$.
At higher weights, the reduction in questions favors exact selection.
Figure~\ref{fig:sensitivity} reweights the same recorded actions for each value of $c$.
Within each bootstrap sample we center the costs on the cheapest displayed policy.

\subsection{Score Aggregation}
\begin{figure*}[tp]
\centering
\includegraphics[width=\textwidth]{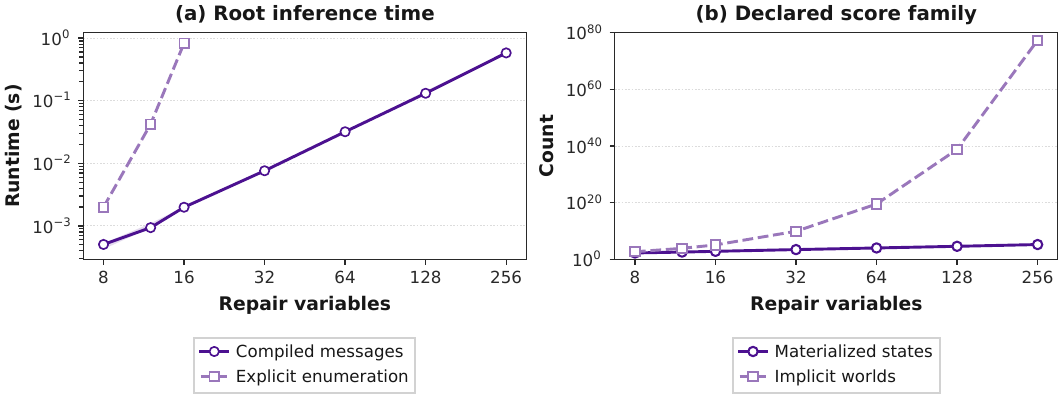}
\caption{Runtime and state counts for score aggregation. Timing curves show the median and observed range over three seeds per setting.}
\label{fig:scaling}
\end{figure*}
For bounded integer scores, forward messages count how many assignments attain each score \citep{todor2013massive}.
A reverse pass recovers the contributions of individual variables using differential inference \citep{darwiche2003differential}.
Figure~\ref{fig:scaling} compares the resulting score histograms with enumeration through 16 repair variables, the largest enumerated setting.
The histograms agree throughout this range.
The computation is pseudo-polynomial in score range. Across three seeds it covers $2^{257}$ worlds in 0.55 to 0.59 seconds at 256 repairs.

\section{AmbiQT Audit Details}
\label{app:audit}
\begin{table*}[!t]
\centering\small
\setlength{\tabcolsep}{5pt}
\begin{tabular}{lrrrrrrr}
\toprule
Model & Outcomes & Valid SQL (\%) & Generated & Kept: obs./two & Tokens & $t_{50}$ (s) & $t_{90}$ (s) \\
\midrule
Doubao Lite & 155 & 97.98 & 3.51 & 1.99 / 2.24 & 2001 & 32.00 & 48.66 \\
GPT Sol & 155 & 98.57 & 2.71 & 1.47 / 1.76 & 575 & 5.02 & 11.11 \\
\bottomrule
\end{tabular}
\caption{Candidate-generation resources and SQL validity, reported once per model.}
\label{tab:model-resources}
\end{table*}

\begin{figure*}[t]
\centering
\includegraphics[width=\textwidth]{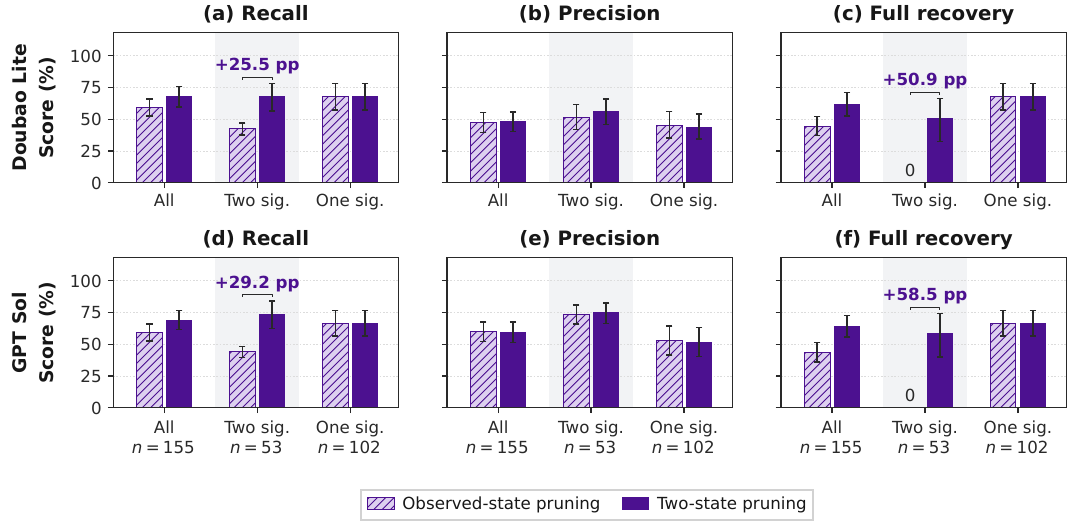}
\caption{Reference recovery by pruning rule. Bars show 95\% database bootstrap intervals.}
\label{fig:support}
\end{figure*}
\begin{figure*}[tp]
\centering
\includegraphics[width=\textwidth]{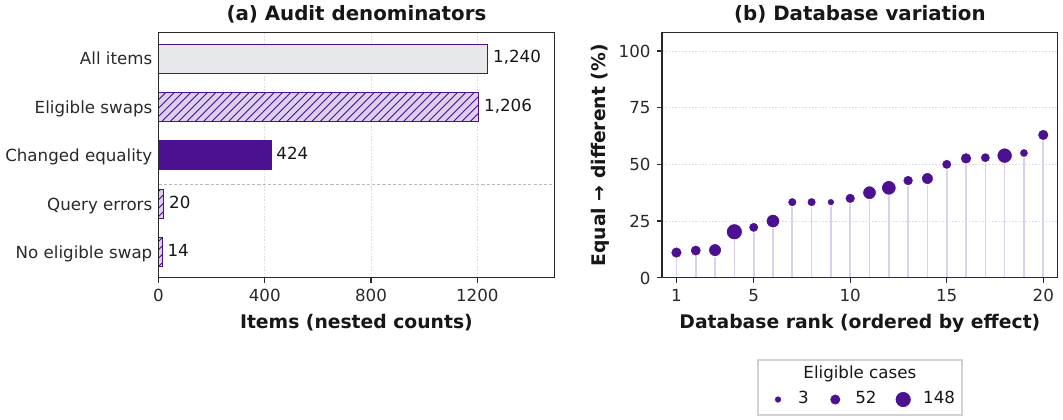}
\caption{Cases eligible for the AmbiQT value swap and cases where it separates the reference queries. Marker area gives the number of eligible cases per database.}
\label{fig:audit}
\end{figure*}
\begin{figure*}[!t]
\centering
\includegraphics[width=\textwidth]{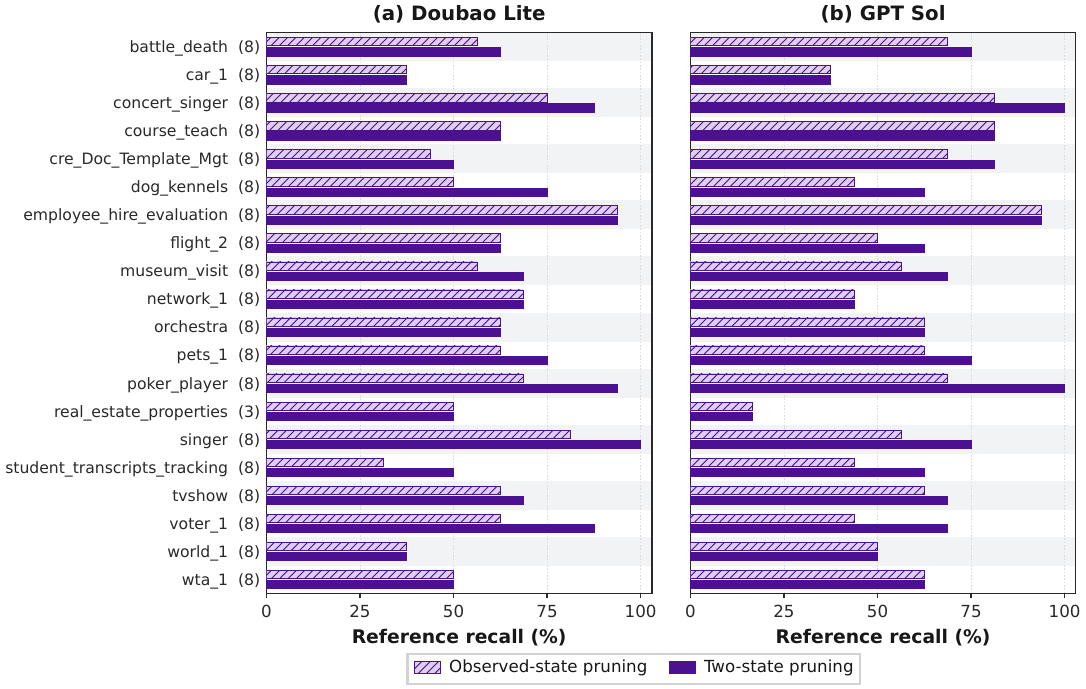}
\caption{Reference recall by database and pruning rule. Parentheses give the number of sampled questions.}
\label{fig:database}
\end{figure*}
To execute the AmbiQT queries, we reconstruct synonym columns from the published mapping.
We preserve SQLite's case-insensitive identifiers and compare text containing invalid UTF-8 as bytes.
We record execution failures separately from valid empty answers.
For eligible cases we choose two rows with distinct non-null values in the second synonym column and exchange those values.
The left panel of Figure~\ref{fig:audit} shows successive stages of case selection.
The right panel orders databases by the fraction of eligible swaps that separate the reference queries.

Figure~\ref{fig:database} shows reference recall and its change under pruning in the 20 sampled databases.
Each value averages over the questions sampled from that database.
The signatures used for scoring contain two outputs.
Different SQL programs can still produce the same signature if their answers agree on both tested states.

Figure~\ref{fig:support} compares recall, precision, and full recovery by reference set size. Table~\ref{tab:model-resources} reports per-question candidate and token counts. Median call times were 32.00 seconds for Doubao Lite and 5.02 seconds for GPT Sol.

\section{Evaluation Protocols}
\label{app:repro}
\paragraph{Tasks And Seeds.}
The sales generator uses seeds 0 to 99 for development and 10000 to 10039 for the policy comparison.
Stress seeds 20000 to 20007 each give six fixtures.
We weight all 16 hidden worlds equally.
Across 120 sales tasks, 48 stress fixtures, and ten configurations, the evaluation contains 26,880 episodes.
The paired bootstrap uses seed 20261029 and 2,000 resamples.
We resample seeds for the sales comparisons and databases for the model comparisons.

\paragraph{Policy Accounting.}
Each previously unseen answer costs one execution, so a full map costs 16.
The evaluator checks returns and question relevance against the complete map.
It checks certificates against the answers the policy knew at the time of each decision.
Question violation rates divide by the number asked and are undefined when a policy asks none.

Cost curves reweight the same recorded actions.
Timings cover one pass of the Python decision logic and answer lookup.
SQLite map construction and auditing occur outside that pass.

\paragraph{Candidate Counts.}
Most generated queries execute on both database states (97.98\% for Doubao Lite and 98.57\% for GPT Sol).
The models generate 3.51 and 2.71 queries per question on average.
Two-state pruning retains 2.24 and 1.76 distinct candidates per question, compared with 1.99 and 1.47 under observed-state pruning (Table~\ref{tab:model-resources}).
Even after preserving these differences, both models recover fewer than 70\% of the reference signatures.
Changing how candidates are grouped can recover only answers present among the generated queries.

\paragraph{Model Generation.}
The API identifiers were \texttt{doubao-seed-2-0-lite-260428} and \texttt{gpt-5.6-sol-2026-07-09}.
Each model produced 155 outputs.
The providers reported 310,128 and 89,063 tokens, respectively.
We measure elapsed time per generation call and count malformed or truncated outputs as failures on the original request.
A generated query counts as executable if it runs successfully on both the original and modified databases.

\paragraph{Input And Output.}
Each request contains the schema and the ambiguous question. The model supplies one to four interpretations in JSON, each with a short meaning and a SQLite query. The completion budget is 4,096 tokens. Reference queries and table rows are withheld. Appendix~\ref{app:candidate-materials} gives the prompt and mismatch coding rules.
We rank each database's items by a SHA256 hash of seed and item ID and keep up to eight distinct questions.

\paragraph{Scoring Rules.}
We run each query in read-only mode on both states, with a four-second execution deadline and a limit of 10,000 result rows. Both executions must succeed for the query to enter the comparison.
For observed-state pruning we keep the first valid query for each distinct observed output, in generation order. Two-state pruning keeps one representative for each distinct pair of outputs.
Let $R$ be the set of reference pairs and $S$ the retained pairs. Recall is $|R\cap S|/|R|$, precision is $|R\cap S|/|S|$ (zero for an empty $S$), and full recovery is the indicator of $R\subseteq S$.
We average each metric over questions. In the paired bootstrap, a sampled database contributes all of its questions to both pruning rules.

\paragraph{Replication Details.}
The repositories record the evaluation protocols, exclusions, request signatures, and analysis scripts. Tables and statistical plots use the saved measurements.

\section{Worked Examples}
\label{app:examples}
\subsection{Revenue Clarification Trace}
Table~\ref{tab:revenue-worlds} lists the four worlds in Figure~\ref{fig:overview}.
Suppose the assistant first executes the all-orders query with the pending amount set to zero, then executes it with the amount set to 50.
The answers differ only because of the amount. The immediate policy can now ask about the missing value.

\begin{table}[t]
\centering
\begin{tabular}{lrr}
\toprule
Interpretation & $r_0$ & $r_{50}$\\
\midrule
All orders & 100 & 150\\
Completed orders & 100 & 100\\
\bottomrule
\end{tabular}
\caption{Revenue in dollars when order \texttt{O-103} has amount 0 or 50. Each column fixes a possible repair.}
\label{tab:revenue-worlds}
\end{table}

If the response is zero, only the first column survives. Executing the completed-orders query in that column gives 100. Both surviving answers are now known, so the policy returns 100 after three executions and one question.
If the response is 50, executing the completed-orders query gives 100 alongside the known answer of 150. These two worlds witness the intent question. Either response selects a world whose answer is already known. Across this branch the user answers twice and the assistant executes three worlds.

\subsection{Partial Boolean Map}
Consider the Boolean question of whether revenue exceeds 100 in Table~\ref{tab:revenue-worlds}. Encode intent $q$ as completed orders (0) or all orders (1), and repair $r$ as amount 0 or 50.
For illustration, suppose the labels for $(q,r)=00,01,11$ are known to be $0,0,1$. The label at $10$ is still unknown.
The worlds $01$ and $11$ give a relevance witness for $q$.
In the four-cell bound, $a=b=0$, $d=1$, and $c\in\{0,1\}$. Hence $16(G(q)-G(r))=c$.
If the missing label is zero, the gains tie. If it is one, $q$ has strictly higher gain. Both completions certify the intent question from three executions.

\subsection{Song Title Query}
\label{app:song}
AmbiQT Candidates includes a question about songs by singers above the average age (column-synonym validation index 16 in \texttt{concert\_singer}).
The first reference query is
\begin{quote}
\small\ttfamily\raggedright
SELECT track\_name\\
FROM singer\\
WHERE age >\\
\quad (SELECT AVG(age) FROM singer)
\end{quote}
The second reference query selects \texttt{composition\_name} with the same predicate.
The columns initially give the same list. Swapping two \texttt{composition\_name} entries changes only the second result.

GPT Sol generated two executable queries for this question. Pruning on the original database kept one reference signature, while two-state pruning kept both.
Doubao Lite generated four executable queries. Two selected titles directly from the singer table. The other two joined \texttt{singer\_in\_concert}, adding a performance requirement absent from the question. The four queries produced three two-state signatures, of which two matched the references.
Its recall rose from $1/2$ to $2/2$ and its precision from $1/2$ to $2/3$.
The extra signature lowers precision.

\section{Candidate Audit Materials}
\label{app:candidate-materials}
\subsection{SQL Candidate Prompt}
Both models received the same single user message. Bracketed fields below stand for the schema without table contents and the question as written in AmbiQT.

\par\smallskip
\noindent\fbox{\begin{minipage}{\dimexpr\columnwidth-2\fboxsep-2\fboxrule\relax}
\small\raggedright
Translate an ambiguous question into the distinct plausible SQLite SELECT queries it can mean. Use the schema exactly. Return JSON only:
{\ttfamily\upshape \{"interpretations":[\{"meaning":"short explanation", "sql":"SELECT ..."\}]\}.}
Include up to four genuinely different plausible readings; do not invent columns or irrelevant interpretations. No data modification. The schema may contain columns with overlapping names.\par
Schema:\par
\textit{[schema]}\par
Question:\par
\textit{[question]}
\end{minipage}}
\par\smallskip

For the case in Appendix~\ref{app:song}, the question was \emph{List all song names by singers above the average age.} GPT Sol supplied two interpretations. The first named tracks, the second compositions. Both SQL queries used the same age predicate. The evaluator executed the \texttt{sql} field of each entry in response order. The \texttt{meaning} field played no part in scoring. Both pruning rules thus received the same candidate queries.

\subsection{Mismatch Coding Rules}
The unit in Table~\ref{tab:sql-errors} is one candidate occurrence before pruning. Two candidates with the same output pair remain separate occurrences. For each sampled candidate, the annotation packet contained the question, the candidate SQL and both reference queries. It withheld model identity, response position, and the model's explanation.

The annotator compared the candidate with the closer reference. A different returned field or column order counted as \emph{projection}. A join that changed the population counted as \emph{relation}, even if its effect was to exclude rows. A change to a selection predicate counted as \emph{filter}. Grouping, counting units, and duplicate handling fell under \emph{aggregation}. Ordering and ties counted as \emph{ranking} when they affected which rows were selected. The protocol reserved \emph{other} for a specific discrepancy outside these categories and \emph{unresolved} for cases without a defensible label.

A candidate that returned both synonym columns when each reference returned just one received a single projection label. In another case, a candidate joined documents through either of two template keys and selected three tied templates where each reference selected one. It received both relation and ranking labels. One candidate can therefore contribute to several category counts.

\end{document}